\documentstyle[12pt,aasms4]{article}

\def\Lsol{L$_\odot$}

\def\lsim{\mathrel{\lower .85ex\hbox{\rlap{$\sim$}\raise
.95ex\hbox{$<$} }}}
\def\gsim{\mathrel{\lower .80ex\hbox{\rlap{$\sim$}\raise
.90ex\hbox{$>$} }}}

\def\vi{(V$-$I)}

\begin{document}

\def\thefootnote{\fnsymbol{footnote}}

\title{Tracing the Outer Structure of the Sagittarius Dwarf Galaxy:
Detections at Angular Distances Between 10 and 34 Degrees
\footnote{Based on observations obtained with the Blanco Telescope
at CTIO, which is operated by the National Optical Astronomy
Observatory, under contract to AURA.}} 

\def\thefootnote{\arabic{footnote}}

\author{Mario Mateo\altaffilmark{1}}
\affil{\tt e-mail: mateo@astro.lsa.umich.edu}
\author{Edward W. Olszewski\altaffilmark{2}}
\affil{\tt e-mail: eolszewski@as.arizona.edu}
\author{Heather L. Morrison\altaffilmark{3}}
\affil{\tt e-mail: heather@vegemite.cwru.edu}

\altaffiltext{1}{Department of Astronomy, University of Michigan, 821
Dennison Bldg., Ann Arbor, MI \ \ 48109--1090}
\altaffiltext{2}{Steward Observatory, 933 N.~Cherry, University of Arizona, Tucson,
AZ \ \ 85721-0065} 
\altaffiltext{3}{Cottrell Scholar of Research Corporation, and NSF Career Fellow;
Department of Astronomy and Department of Physics, 
Case Western Reserve University, Cleveland OH \ \ 44106}

\begin{abstract}

We have obtained deep photometric data in 24 fields along the
southeast extension of the major axis of the Sagittarius dwarf
spheroidal (Sgr dSph) galaxy, and in four fields along the northwest
extension.  Using star counts at the expected position of the Sgr
upper main-sequence within the resulting color-magnitude diagrams
(CMDs), we unambiguously detect Sgr stars in the southeast over the
range 10--34$^\circ$ from the galaxy's center.  If Sgr is symmetric,
this implies a true major-axis diameter of at least 68$^\circ$, or
nearly 30 kpc if all portions of Sgr are equally distant from the
Sun. Star counts parallel to the galaxy's minor-axis reveal that Sgr
remains quite broad far from its center.  This suggests that the outer
portions of Sgr resemble a stream rather than an extension of the
ellipsoidal inner regions of the galaxy.  The inferred V-band surface
brightness (SB) profile ranges from 27.3-30.5 mag arcsec$^{-2}$ over
this radial range and exhibits a change in slope $\sim 20^\circ$ from
the center of Sgr.  The scale length of the outer SB profile is
17.2$^\circ$, compared to 4.7$^\circ$ in the central region of Sgr.
We speculate that this break in the SB profile represents a transition
from the main body of Sgr to a more extended `Sgr stream'.  By
integrating the SB profile we estimate that the absolute visual
magnitude of Sgr lies in the range $-13.4$ to $-14.6$, depending on
the assumed structure of Sgr; an upper limit to the luminosity of Sgr
is therefore $L \sim 5.8 \times 10^7$ \Lsol.  This result lowers the
M/L$_V$ ratio inferred for Sgr by Ibata et al. (1997) down to
$\sim$10, consistent with values observed in the most luminous dSph
companions of the Milky Way.

\vskip1em

\hang\hang\noindent Subject headings: Galaxy: halo --- Galaxy: structure ---
galaxies: individual (Sgr dSph) --- 
galaxies: interactions --- galaxies: Local Group

\end{abstract}

\clearpage

\section{Introduction}

Since its discovery in 1994 (Ibata et al.), the known extent of the
Sagittarius dwarf galaxy (Sgr) has grown steadily.  The first map
revealed a galaxy with a possibly clumpy structure subtending a region
about $8^\circ \times 5^\circ$ in size, and oriented roughly
perpendicular to the Galactic plane.  Subsequently, a series of papers
reported a far larger projected size of at least $20^\circ \times
8^\circ$ for Sgr (Mateo et al. 1996; Alard 1996; Fahlman et al. 1996;
Ibata et al. 1997; Alcock et al. 1997) based on observations of
various stellar tracers. Recently, Siegel et al. (1997) reported a
possible detection of Sgr far beyond this $20^\circ \times 8^\circ$
boundary.

Because Sgr is located only 16 kpc from the center of the Milky Way
(Ibata et al 1994; Mateo et al. 1995), there is universal agreement
that it is experiencing a strong tidal encounter. Generic simulations
of dwarf-satellite destruction, as well as models specific to Sgr
(Allen and Richstone 1988; Moore and Davis 1994; Piatek and Pryor
1995; Oh et al. 1995; Johnston et al., 1995; Velazquez and White 1995;
Ibata et al. 1997; Zhao 1998) reveal that a strong tidal encounter
will draw leading and trailing tidal streams out from the main body of
Sgr during its closest encounters with the Milky Way.  These streams
should extend close to the projected major axis of the galaxy as stars
migrate along orbital paths close to that of the disintegrating system.
Evidence already exists for extra-tidal stars in some halo systems
(globulars, Grillmair et al. 1995; dSph's, Kuhn et al. 1996, see
discussion in Olszewski 1998).  In this {\it Letter} we report the
systematic search for and discovery of a distinct tidal stream of
stars associated with Sgr but extending nearly three times further
from the center of Sgr than previously published.

\section{Observations and Reductions}

We obtained data on two photometric nights (July 1/2 and 2/3, 1998)
with the BTC CCD array at the prime focus of the Blanco 4m telescope
at CTIO.  Because of charge-transfer problems with Chip~1 of the BTC,
we did not use results from this CCD in our analysis; thus, each CMD
represents counts from 0.19 deg$^2$ on the sky.  We generally obtained
single exposures in V and I (typically 10-12 min duration each) in
every field.  An automatic pipeline procedure (Unix and IRAF daemons)
was developed to process the frames with no intervention and in near
real time.  Data were archived, overscan, zero and flatfield
corrected, reduced using the DoPhot photometry program, and combined
to make instrumental CMDs.  Results for three Sgr fields and their
control fields are shown in Figure~2.  The CCDs were individually
calibrated using numerous observations of Landolt (1992) standards.
The photometric transformations for I and \vi\ showed a scatter of
less than 1.5\%\ on both nights.

Our goal was to identify main-sequence stars in Sgr along the
extension of its major axis, the position angle (PA) of which we originally
took to be $109.2^\circ$ (we present a better estimate of this angle
below).  The center of Sgr was assumed to coincide with the globular
cluster M~54.  The locations of the 17 major-axis fields to the
southeast are shown in Figure~1 and their equatorial and Galactic
coordinates are listed in Table~1.  We also observed seven control
fields located symmetrically opposite of the $l = 0^\circ$ meridian
from the corresponding target fields (see Figure~1 and Table~1), and
four fields perpendicular to the major axis of Sgr and passing through
Field~3 (denoted as Fields 3-S$n$ and 3-N$n$ for the fields south and
north of Field~3, respectively, in Table~1).  The angular separation
of each field from the center of Sgr is denoted as $R$ in Table~1.

\section{Analysis and Results}

The Sgr main sequence turnoff occurs at I$_0 \sim 20$ mag and \vi\
$\sim 0.6$ (Mateo et al. 1996; Fahlman et al. 1996; Figure~2).  By
good fortune, this corresponds to a gap in the density of
contaminating foreground stars from the thin disk (located at I$_0
\lsim 20$ in this color range) and background galaxies that populate
the prominent `blob' in Figure~2 fainter than $I_0 \sim 22.5$ at this
color.  To improve contrast, we counted stars only within the box
shown in the first panel of Figure~2.  Subtracting the counts from a
control field statistically removes non-Sgr stars in the box.  The
reddening at every field center was taken from the extinction maps of
Schlegel et al. (1998; see Table~1); we shifted the selection box
accordingly before counting stars.  At the lowest Galactic latitudes
the 3-chip BTC area contains about $10^4$ stars; typical separations
of all detected objects are therefore 10 times the seeing
diameter. From past experience (e.g. Hurley-Keller et al. 1998), we
conclude that completeness corrections are negligible in all of our
fields.

The upper panel of Figure~3 shows the main-sequence counts along the
Sgr major-axis and for the control fields.  The effects of systematic
changes in the adopted \vi\ reddening by $\pm$0.03 mag are also
illustrated.  Even in the outermost Sgr point -- Field 17 located at
$R = 34^\circ$ -- there is a clear excess of stars relative to the
controls.  Note that the {\it raw} Sgr counts exhibit a clear change
in the slope of the density profile at $R \sim 20^\circ$.

To correct for contamination by Galactic stars, we fit a straight line
to the control-field counts (Fig 3), subtracting the fitted counts
from the raw Sgr counts.  The net counts are plotted as a function of
$R$ in the lower panel of Figure~3.  Mateo et al. (1996) estimated
that the V-band surface brightness (SB) at the innermost field in this
study ($R \sim 10^\circ$) is $\Sigma_V = 27.3$.  Normalizing to this
value, we can express the star counts in SB units as shown in the
lower panel of Figure~3.  We assume that there are no spatial
variations in the stellar population within Sgr, a point supported by
star counts in other regions of our CMDs. The SB of the outermost Sgr
fields, where we detect a 4-6$\sigma$ excess over the control field
counts, is $\Sigma_V \sim 30.5$ mag arcsec$^{-2}$.

Since a tidal stream should consist of leading and trailing
components, we also observed fields at 30$^\circ$ and 40$^\circ$ along
the northwest major axis extension.  However, the large and rapidly
variable reddening at (l,b)$=$ (3,16) and (2,26) caused this first
attempt to fail.  Siegel et al.~(1997) also failed to find Sgr more
than 40$^\circ$ towards the northwest at (l,b)$=$(353,41).

\section{Discussion}

Given the clear break in the star-count profile in Figure 3, we have
fit the net counts with two exponential profiles.  The
best-fit composite profile is plotted in Figure~3 and can be
decomposed as:
$$\Sigma_{V,inner} = 25.25 + 0.23 R,$$
$$\Sigma_{V,outer} = 28.55 + 0.063 R,$$ where $\Sigma$ is in mag
arcsec$^{-2}$, and $R$ in degrees.  Due to time lost to poor weather,
our data perpendicular to the major axis are limited so far to the one
cross-cut centered on Field 3 (see Figure~1).  Taking $z$ to be
the angular separation from the major axis along this cross-cut (in
degrees), we find that a Gaussian profile adequately fits the
data:
$$\Sigma_{cc,3} = 27.84 + 0.071 z^2.$$ The constant reflects the fact
that the centroid of $\Sigma(z)$ relation is about 1$^\circ$ north of
Field 3.  Either we did not precisely follow the `true' major axis of
Sgr, or else the galaxy's projected stellar density distribution does
not follow a great circle.  In the former case, we conclude that the
true major-axis PA of Sgr is $104.8^\circ \pm 1.2^\circ$.  To test for
curvature, more cross-cuts are required. The Siegel et al.~(1997)
field is $\sim$2$^\circ$ north of our Field 12. Their detection of Sgr
is consistent with the smaller PA, and with the broad width of Sgr at
this location.  Despite Field 17 being $\sim$2.5$^\circ$ away from the
ideal position, it still unambiguously shows Sgr, 7.5$^\circ$ further
from the center of Sgr than the Siegel et al. field.

If we extrapolate the composite exponential profile in Figure~3 to the
center of Sgr, the inferred central SB is $\Sigma_{0,V} = 25.2 \pm
0.3$ mag arcsec$^{-2}$, close to the value estimated by Mateo et
al. (1995).  Ibata et al. (1997) found that the radial profile of Sgr
is well fit by a King-profile with a core radius of $\gsim
1.3^\circ$. Since exponential and King-model profiles fit
low-concentration systems equally well (e.g., Eskridge 1988), these
results are not incompatible.  We also note that the smooth
exponential profiles do not seem to perfectly match the observations
on 3-5$^\circ$ scales (Figure~3; note the Poisson error bars).  This
implies that the SB distribution of Sgr may be clumpy in its outer
regions (Mateo et al. 1996); further data are needed to confirm this.

We consider two possible cases to estimate the total luminosity of
Sgr.  First, the inner exponential is taken to represent the `main
body' of Sgr with a 3:1 axis ratio (Ibata et al. 1997), while the
outer profile represents a tidal stream of constant width as defined
by the Guassian $\Sigma(z)$ relation above.  By integrating these
profiles, we derive an integrated apparent magnitude of $V_{tot,1} =
3.6$ for Sgr.  If the distance modulus over the entire extent of the
galaxy is assumed constant at $(m-M)_0 = 17.0$ (Ibata et al. 1994;
Mateo et al. 1995; Ibata et al. 1997; though see Alcock et al. 1997),
then $M_{V,tot,1} = -13.4 \pm 0.3$.  A second approach is to assume
that the width of Sgr is constant at all values of $R$, again obeying
the $\Sigma(z)$ relation derived above.  For this case, $V_{tot,2} =
2.7 \pm 0.3$, and $M_{V,tot,2} = -14.3$.  Previous estimates of the
integrated absolute magnitude of Sgr range from $-13$ to $-13.5$
(Ibata et al. 1994; Mateo et al. 1996; Ibata et al. 1997).  If we
correct for the slight offset of our major axis from the `true' major
axis (which slightly increases the radial scale lengths), we can raise
the integrated luminosity of Sgr to at most $M_{V,tot,1} \geq -13.8$
and $M_{V,tot,2} \geq -14.6$.  Unless an additional outer component of
Sgr remains to be discovered, it seems very unlikely that the
integrated V-band absolute magnitude of Sgr is brighter than $-14.6$,
corresponding to $L_{Sgr,tot} \leq 5.8 \times 10^7$ \Lsol.

An important possible consequence of this result is that the inferred
V-band M/L ratio of Sgr could drop from $\sim$50 (Ibata et al. 1997)
to a value as low as $\sim$10 with our larger estimate of the total
luminosity of Sgr.  This lower mass-to-light ratio brings Sgr into
better consistency with the M/L ratios observed in other dSph
satellites of the Milky Way and M31 (Bellazzini et al. 1996; Mateo
1998).

Perhaps the most interesting feature of the radial profile of Sgr is
the kink at $R \sim 20^\circ$.  Does this represent the transition
from a dynamically distinct portion of the Sgr dwarf and a tidal
stream -- the `Sgr stream' -- pulled out of the galaxy?  Or are we
seeing the current (inner) tidal stream in the process of joining an
older, more extended stream from a previous tidal encounter (Zhao
1998)?  One key to addressing these questions will be to define the
orbital path of Sgr by determining distances along the stream, as
attempted by Alcock et al. (1997), and to better define the projected
distribution of Sgr stars on the sky.  The only well-populated
features in the Sgr CMD in these outer fields are the main sequence
and sub-giant branch, neither of which are well suited for precise
distance determinations.  Nevertheless, we have unambiguously detected
Sgr at an angular distance that implies overall dimensions of at least
$68^\circ \times 8$-$10^\circ$, or about $30 \times 4$ kpc.  Since the
projected scale length of the outer parts of Sgr is 17.2$^\circ$, we
should be able to use the techniques described in this paper to trace
the galaxy 20-30$^\circ$ further out than we have mapped it so far.
Wide-field time series observations along the stream may reveal RR~Lyr
variables or dwarf Cepheids that can be used to determine precise
distances along the outer extension of the Sgr dwarf.

\vskip3em

\section{Acknowledgements}

The authors have been supported by NSF grants AST~9619490 and
AST~9619524.  We thank Mauricio Fernandez for his help at the 4m
telescope, and the entire CTIO staff for their good cheer during
abyssmal -- at times, apocalyptic -- weather over the past three
observing seasons.  We thank Gary Bernstein, Denise Hurley-Keller,
Craig Kulesa and Tim Pickering for help in producing Figure~1.

\clearpage

\centerline{\bf Figure Captions}
\vskip2em

\noindent{\bf Figure 1} -- A chart showing the locations of the Sgr
(open squares) and control fields (open circles).  Also noted are the
Sgr globular clusters (M~54, Ter~7, Ter~8 and Arp~2), and the
foreground cluster M55 (see Mateo et al. 1996; Fahlman et al. 1996).
We also plot the approximate extent of Sgr from Ibata et al. (1997;
small ellipse denoted `IGI'), and where Alard (1996; large rectangle
denoted `A') and the Macho team (Alcock et al. 1997; rectangle denoted
`Macho') found RR~Lyr stars. The large dashed ellipse is the
approximate extent of Sgr after these studies and the work of Ibata et
al. (1997).  The short dashed line corresponds to the SB break seen in
Figure~3.

\vskip1em

\noindent{\bf Figure 2} -- Dereddened color-magnitude diagrams of Sgr
Fields 3, 10 and 16 (upper panels; $R$ is labeled in each panel; see
Table~1), and the corresponding three control fields (lower panel).
The upper-left panel shows the main-sequence box used to count stars.
The extinction parameters were taken from Schlegel et al. (1998).

\vskip1em

\noindent{\bf Figure 3} -- {\it Upper panel:} The raw star counts in
the main-sequence box (see Figure~2) as a function of $R$ for both the
major-axis Sgr fields (filled squares), and the control fields (filled
circles).  A linear fit to the control data is shown as a dashed line.
The effects of systematically increasing the field reddenings by 0.03
mag (in \vi) is shown by the upper dash associated with each point,
while the lower dashes show the effects of lowering the reddening by
0.03 mag.  {\it Lower panel:} The net counts in each Sgr major-axis
field as a function of $R$ (angular distance from M54).  The large
dashes above and below each data point have the same meaning as in the
upper panel. {\it The conventional error bars denote the Poisson
uncertainties of each point.}  The dotted line represents the
composite exponential profile described in Section 4.  The raw and net counts
are listed in Table~1.

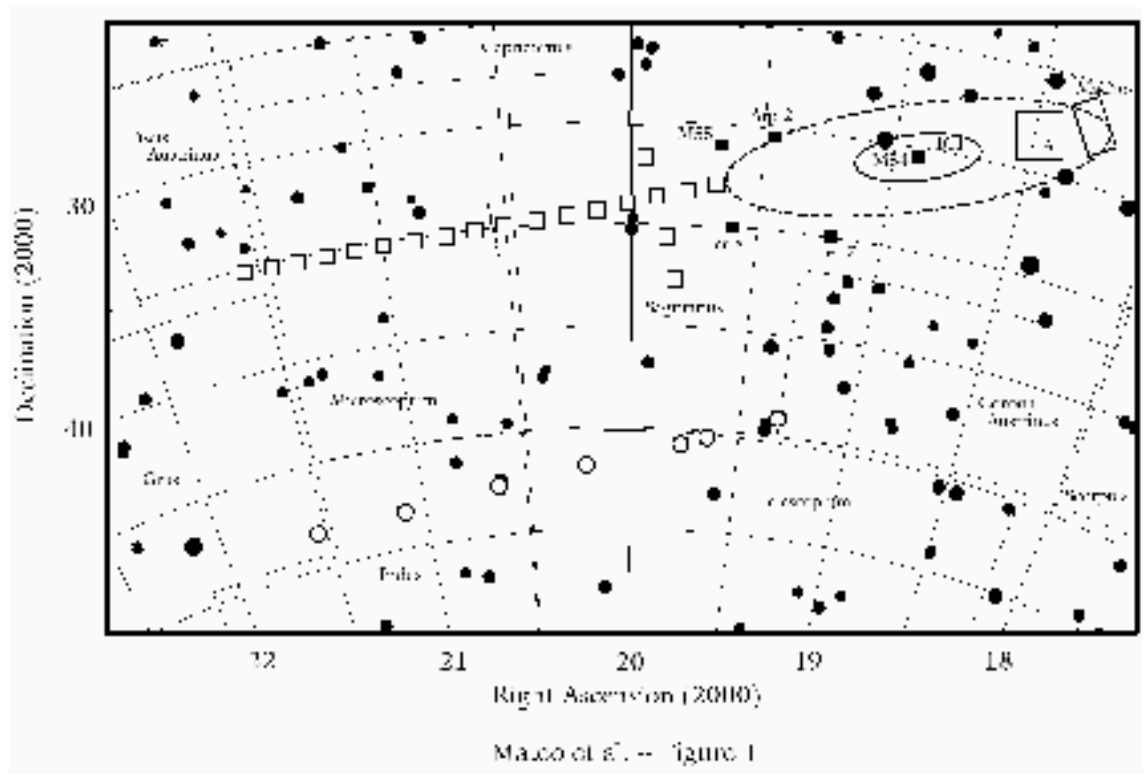
\end{document}